\documentclass[journal]{IEEEtran}
\usepackage[export]{adjustbox}
\usepackage{graphicx}
\usepackage{array}
\usepackage{cite}
\usepackage{color}
\usepackage{amsmath}
\usepackage{url}
\usepackage{float}
\usepackage{algorithm}
\usepackage{tabularx}
\usepackage{algpseudocode}
\usepackage{threeparttable}
\newcolumntype{M}[1] {>{\centering\arraybackslash} m{#1} }
\usepackage{tikz}

\usepackage[font=small]{caption}
\ifCLASSINFOpdf
 
\else
 
\fi
\usepackage{subcaption}
\newcommand{\RN}[1]{%
  \textup{\uppercase\expandafter{\romannumeral#1}}%
}
\usepackage{multirow}
\begin{document}

\title{Design Space Exploration of Power Delivery For Advanced Packaging Technologies}

\author{Md~Obaidul~Hossen,~\IEEEmembership{Student~Member,~IEEE,}
        Yang Zhang,~\IEEEmembership{Student~Member,~IEEE,} Hesam Fathi Moghadam,~Yue Zhang,~Michael Dayringer,
        and~Muhannad S Bakir,~\IEEEmembership{Senior~Member,~IEEE}% <-this % stops a space
\thanks{M. O. Hossen, Y. Zhang, and M. S. Bakir are with the School of Electrical and Computer Engineering, Georgia Institute of Technology, Atlanta,
GA, 30332 USA (e-mail: mhossen3@gatech.edu).}% <-this % stops a space
\thanks{ H. Fathi Moghadam, Y. Zhang, and M. Dayringer are with Oracle Labs, Redwood Shores, CA.}% <-this % stops a space
}

%

% The paper headers
\markboth{Journal of \LaTeX\ Class Files,~Vol.~XX, No.~X, Month~Year}%
{Shell \MakeLowercase{\textit{et al.}}: Bare Demo of IEEEtran.cls for IEEE Journals}

% make the title area
\maketitle

\begin{abstract}
In this paper, a design space exploration of power delivery networks is performed for multi-chip 2.5-D and 3-D IC technologies. The focus of the paper is the effective placement of the voltage regulator modules (VRMs) for power supply noise (PSN) suppression. Multiple on-package VRM configurations have been analyzed and compared. Additionally, 3D IC chip-on-VRM and backside-of-the-package VRM configurations are studied. From the PSN perspective, the 3D IC chip-on-VRM case suppresses the PSN the most even with high current density hotspots. The paper also studies the impact of different parameters such as VRM-chip distance on the package, on-chip decoupling capacitor density, etc. on the PSN. 
\end{abstract}


\IEEEpeerreviewmaketitle

\section{Introduction}

\IEEEPARstart{P}{ower} requirements in modern high performance computing systems are becoming increasingly stringent. Traditionally, the power supplies are placed off-chip to provide necessary load currents to the on-chip active circuitry. These systems typically have resistive and parasitic losses from the interconnects and metal pads. Large passive components (i.e., capacitors) are placed to somewhat compensate these effects. However, the power delivery challenges are becoming increasingly prominent as more and more transistors are being packed into a single chip. As a result, number of bumps/pads, on-chip wires, etc. is also increasing. These added components contribute to the parasitics of the power delivery path. Recently, on-chip regulators have gained significant attention because of their fine grain voltage control, increased availability of power, increased performance, decreased inductor size, etc. \cite{intel:fivr,shepard:ivr,harvard:dvfs,intel:ivr_isscc17}. These technologies eliminate the need for multiple VRs in the case of multiple supply voltage systems while reducing the parasitic length of the power delivery path, enabling active power management required by high performance computing devices. In short, these are efforts to bring the power supply circuitry closer to the active circuits.

Power supply noise (PSN) modeling has been under extensive research over the last couple of decades \cite{jianyong:ddm,voltspot,yang:edl,harvard:vnscope,opu:ectc2018}. The power supply is assumed ideal in the prior studies. Moreover, there are VRM parasitics which impact the PSN of the system. Also, a distributed package level PDN along with the distributed decoupling capacitors are important in accurately characterizing the PSN of any given architecture. Therefore, in this paper, based on prior Power Delivery Network (PDN) modeling efforts \cite{yang:edl, opu:ectc2018, opu:epeps2018}, different voltage regulator module (VRM) placement methodologies e.g., on-package, 3D stacked VRM-chip, VRM placed on the backside of the package, etc. have been explored. First, the benchmark architectures for analysis are described in Section~\RN{2}. Section~\RN{3} shows the DC IR drop results for different architectures. Transient noise of different configurations is analyzed and compared in Section~\RN{4}. Finally, in Section~\RN{5}, the concluding remarks are stated.

\section{PDN Modeling and Specifications}
Several benchmark configurations have been analyzed in this paper.
\begin{figure*}[!ht]
    \centering
    \begin{subfigure}{0.25\textwidth}
        \includegraphics[width=\textwidth]{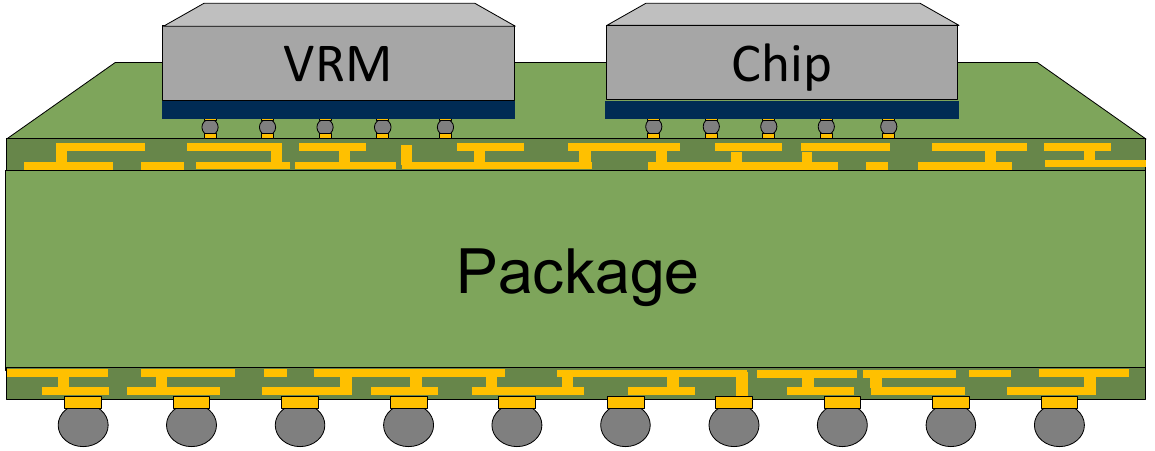}
        \caption{}
        \label{fig:on_pkg_VRM}
    \end{subfigure}% blank line will push the figure to a new line
    ~
    \begin{subfigure}{0.25\textwidth}
        \includegraphics[width=\textwidth]{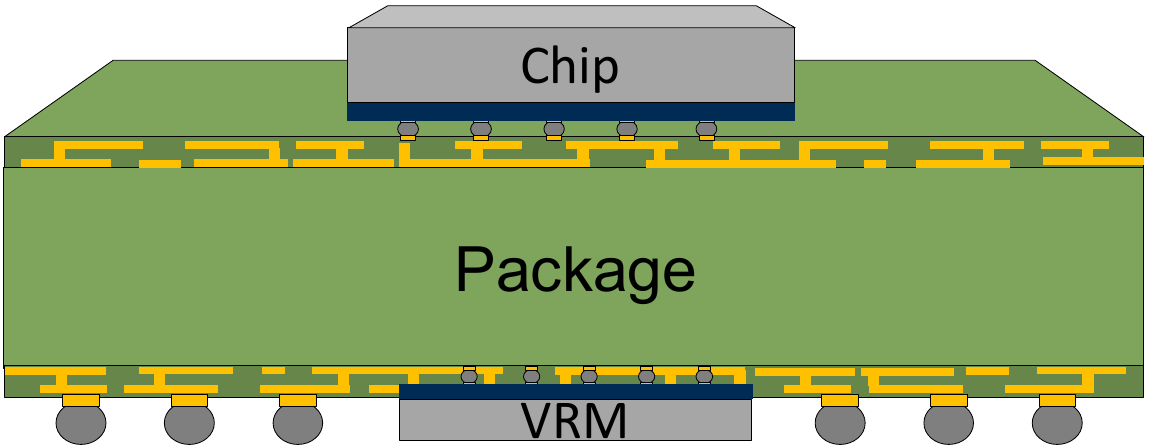}
        \caption{}
        \label{fig:backside_VRM}
    \end{subfigure}
    ~
    \begin{subfigure}{0.25\textwidth}
        \includegraphics[width=\textwidth]{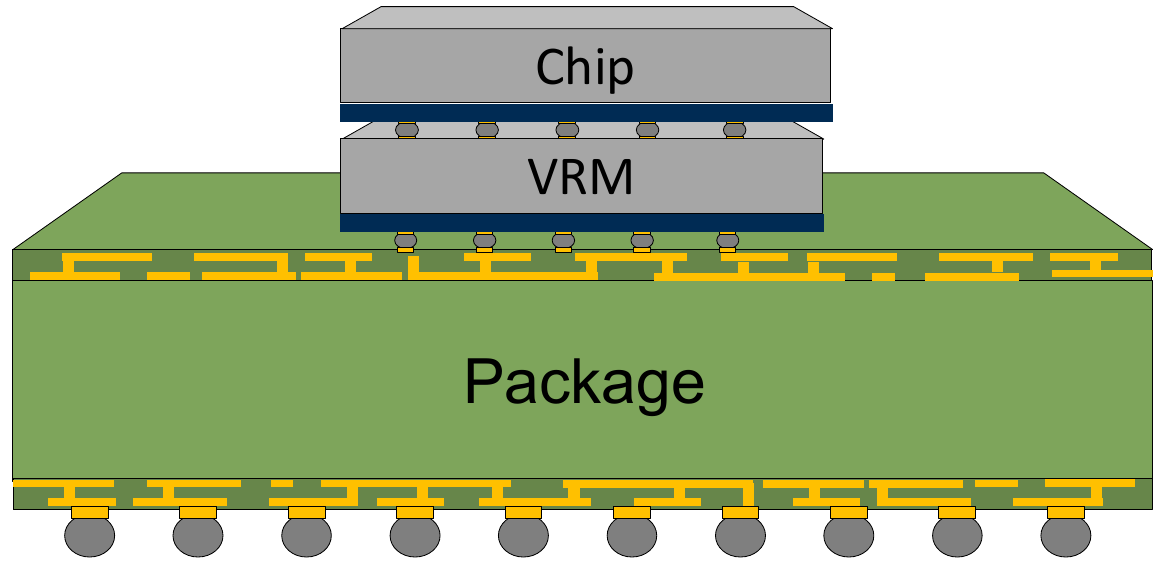}
        \caption{}
        \label{fig:3D_VRM}
    \end{subfigure}
	\caption{Benchmark architectures: (a) On-package VRM configuration, (b) Backside-of-the-package VRM Configuration, and (c) 3D IC Chip-on-VRM Configuration}
%\vspace{-20pt}
\end{figure*}
In Fig. \ref{fig:on_pkg_VRM}, the VRM chip is placed next to the active chip on the same package and thus, the long interconnect distance from the power supply to the chip is reduced. The configuration shown in Fig. \ref{fig:backside_VRM} considers a VRM chip placed on the backside of the package. Fig. \ref{fig:3D_VRM} shows the 3D IC stacking of a processor chip on top of the VRM chip. 
%

%\subsection{PDN Topology and Modeling Methods}
\begin{table}[!ht]
\begin{center}
\caption{PDN parameters}
\begin{tabular}{p{0.2\textwidth} p{0.2\textwidth}}
\hline
\hline
TSV resistivity \cite{ASE:tsv_phro} & 80$\times$\(10^{-9}\) $\Omega$m\\
Package wire thickness (metal planes) & 10 P/G metal layers, 0.010 mm per layer\\
On-chip PDN wire dimensions & 5 um thick, 3.3 um wide, 30 um pitch\\
On-chip PDN wire resistivity & 17.1$\times$\(10^{-9}\) $\Omega$m\\
On-chip decoupling capacitor & 5.3 nF/mm$^2$\\
C4 bump diameter/pitch & 40 $\mu$m/100 $\mu$m\\
\hline
\label{tab:specs}
\end{tabular}
\end{center}

\end{table}

\begin{figure}[!ht]
    \centering
        \includegraphics[width=0.35\textwidth]{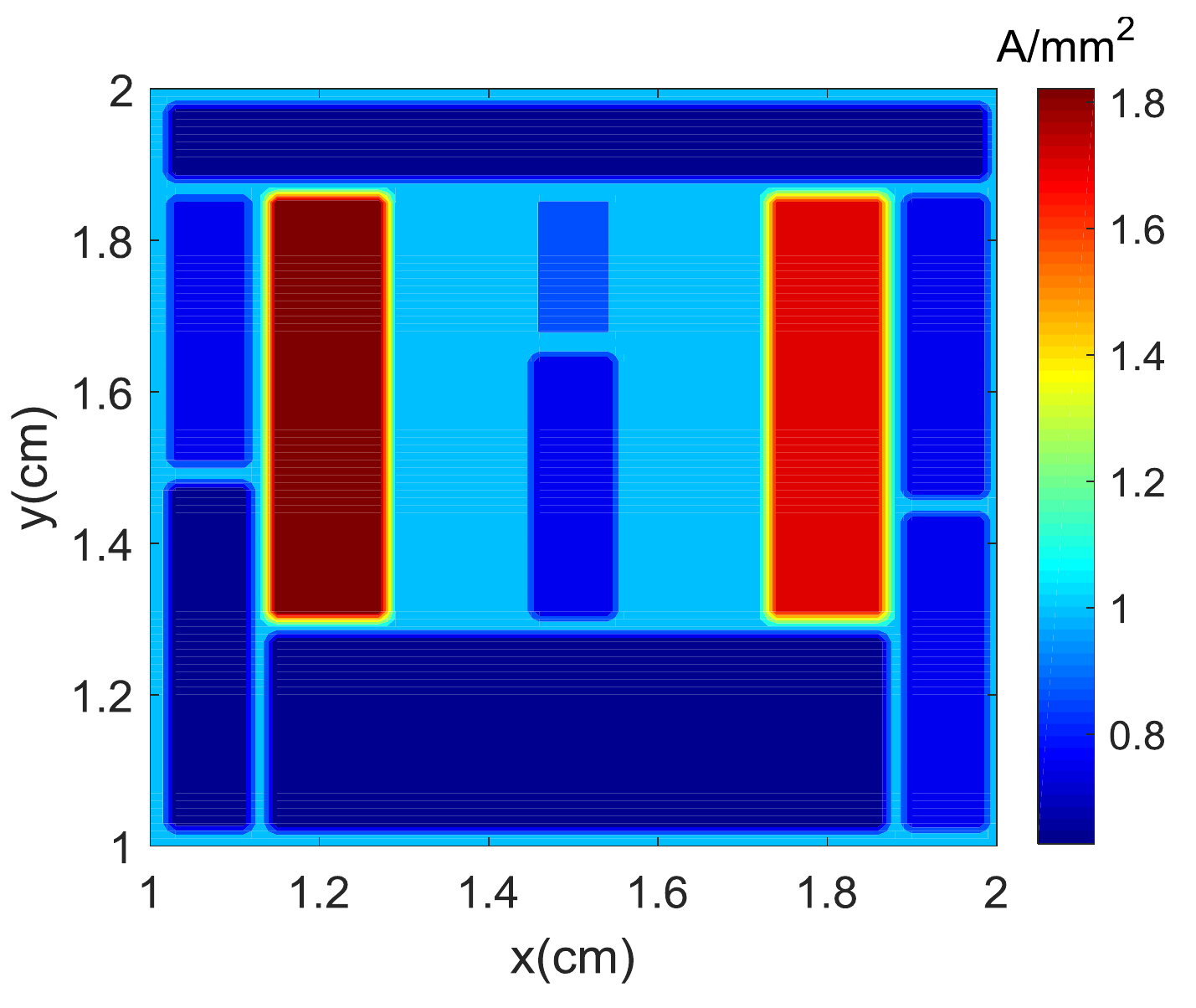}
        
\caption{The non-uniform current density map used for the analysis}
\label{fig:power_map}
\end{figure}

The overall analysis flow is as described in the prior work \cite{yang:edl,opu:ectc2018}. Throughout this paper, a 1~cm $\times$ 1~cm chip is considered. The active chip is assumed to have a 1~V supply voltage rail and a total power of 100 W. VRM parasitic resistance and inductance are extracted from the literature \cite{altera:pdn_tool,onsemi:vrm}. The equivalent series resistance (C$\_$esr) and inductance (C$\_$esl) of the capacitors are also incorporated in the model. The second region is the board-level lumped parameters along with their decoupling capacitance. Bump parasitics have two sources: solder bumps between the package and the board, and C4 bumps between the package and the chip. Discrete decoupling capacitors have equivalent series resistance (ESR), equivalent series inductance (ESL), and equivalent series capacitance (ESC). These decoupling capacitors are then distributed throughout the package. The overall specifications of different parameters are described in Table~\ref{tab:specs}. Non-uniform current density map with distinct high power blocks is used for the simulations. The power map (or current density map) is as specified in \cite{yang:tcpmt1}, but is modified according to \cite{ucsd_hp:multicore, Esmaeilzadeh:multicore}.

%%%%%%%%%% INFORMATION %%%%%%%%%%
\section{DC IR Drop Comparison of Different Benchmark Configurations}
\begin{figure}[!ht]
\centering \includegraphics[width=0.4\textwidth]{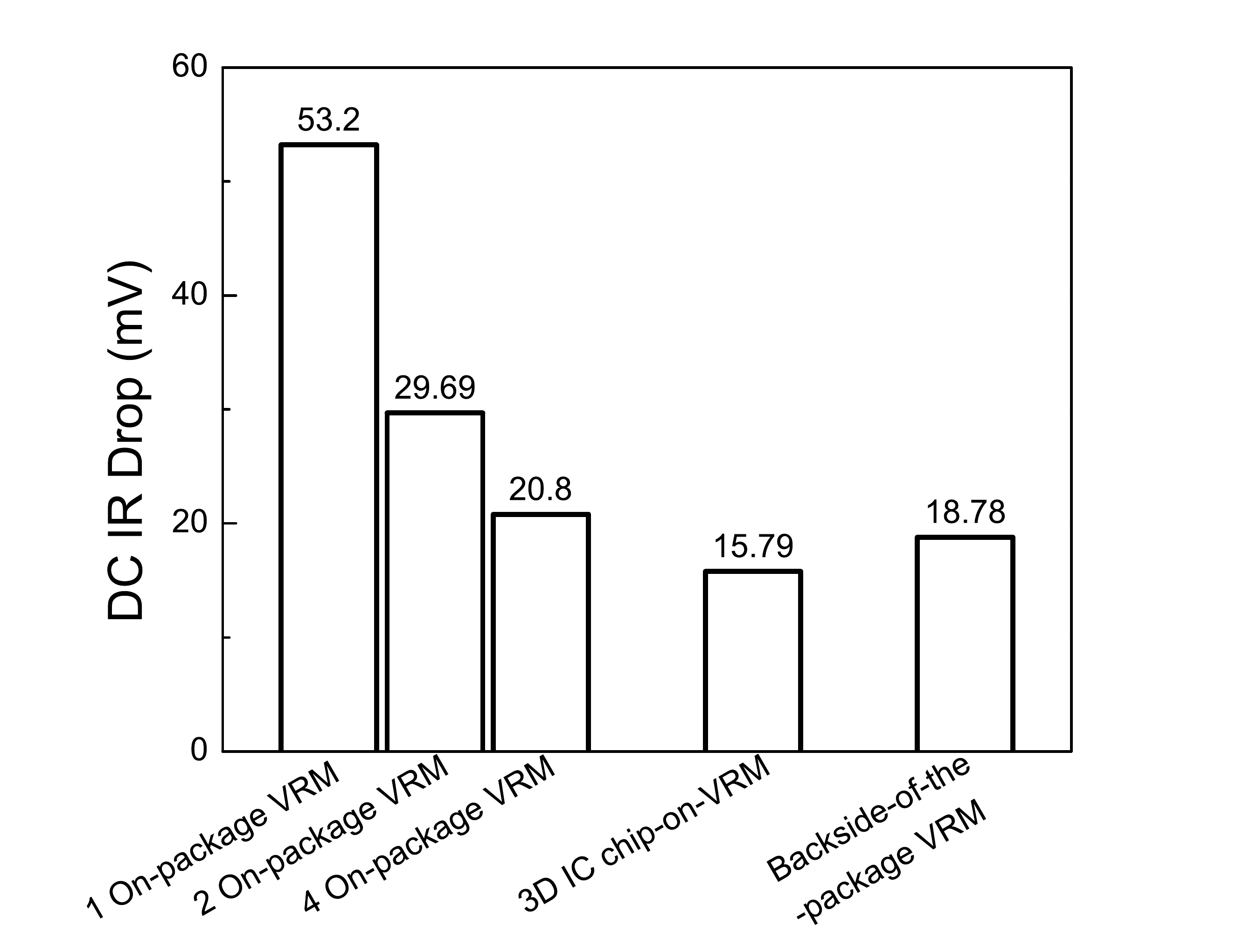}
\caption{Comparison of DC IR drop for different configurations}
\label{fig:pkgVRM_vs_pVRM}
\end{figure}

In this section, the DC IR drop for different configurations i.e., on-package VRM, 3D IC chip-on-VRM, and the backside-of-the-package VRM, etc. have been analyzed. In each configuration, adding additional VRMs to the system reduces  DC IR drop. The impact of multiple VRMs on PSN suppression is more pronounced if there are hotspots in the chip. 
Fig. \ref{fig:pkgVRM_vs_pVRM} summarizes the results for different VRM-processor configurations. In the backside-of-the-package VRM and 3D IC chip-on-VRM cases, owing to the shorter distance, the IR drop is smaller compared to the prior on-package VRM cases. In the backside-of-the-package configuration, the through package vias and metal layers in the package PDN are important components of the power delivery path. In the 3D IC case, however, due to the dense bumps between the chips, the TSVs in the VRM chip and the microbumps between the VRM and the active chip are the only contributors of the parasitics in the PDN path. As a result, the IR drop for the 3D IC case is 24\% and 15.9\% smaller than that with four on-package VRMs, and backside of the package VRM cases, respectively.

\begin{figure}[!ht]
\centering \includegraphics[width=0.4\textwidth]{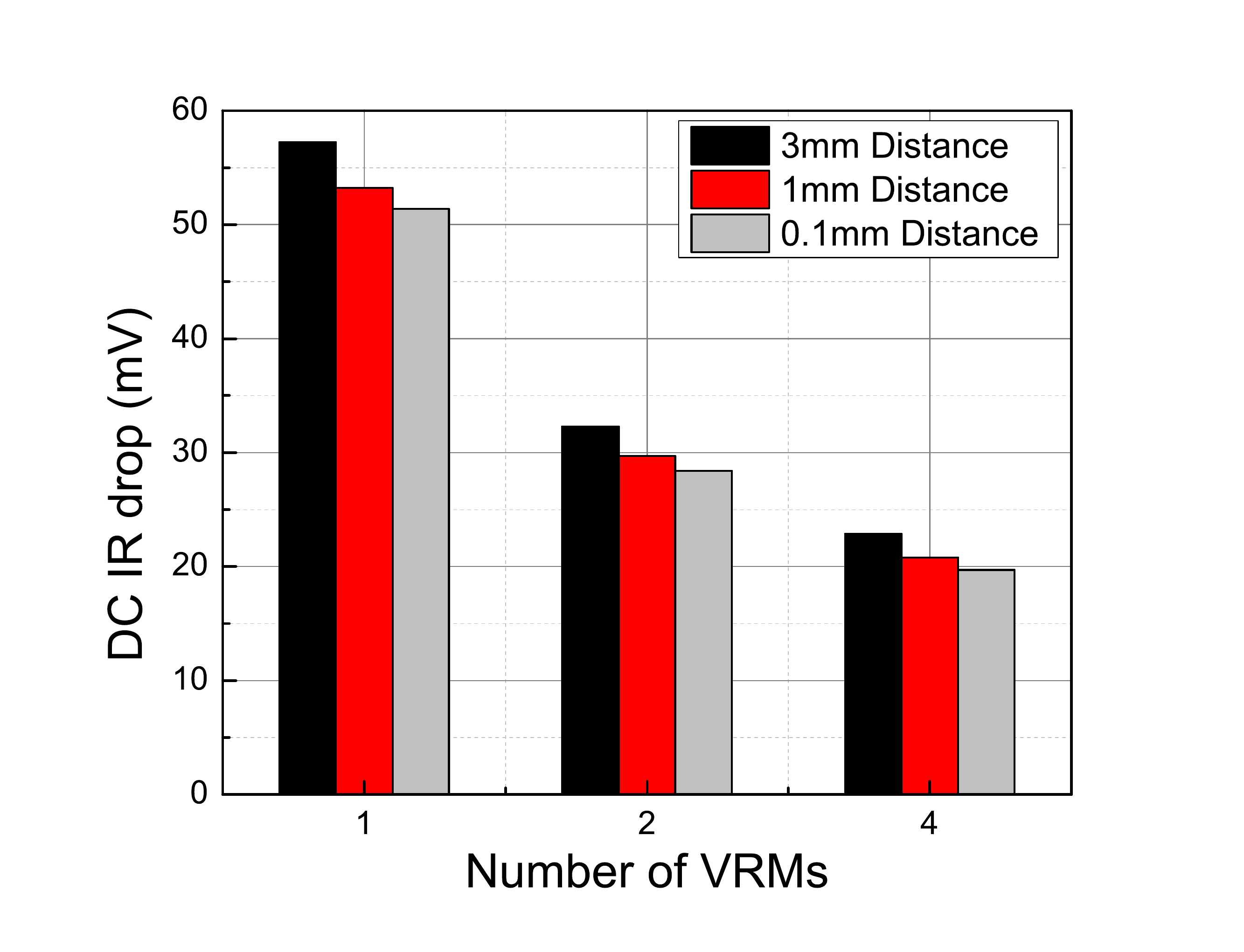}
\caption{Comparison of maximum IR drop for different VRM-chip gaps in the on-package VRM configurations}
\label{fig:gap}
\end{figure}

Since the multiple on-package VRMs case brings the regulator circuit closer to the active chip, there are different trade-off analyses which determine how close we can bring these chips. Fig. \ref{fig:gap} shows the DC IR drop results for three different distances between the chips.
In the baseline model, the distance between the VRM and the processor chip was fixed to 1~mm. To investigate the impact of this on power delivery performance, the distance was varied from 3~mm to 0.1~$m$m. Fig. \ref{fig:gap} summarizes the results for 3~mm, 1~mm and 0.1~mm distances. As expected, if the distance is increased, the interconnect length for power supply increases, which eventually increases the IR drop.
\iffalse
\begin{figure}[!ht]
\centering \includegraphics[width=0.4\textwidth]{figs/bump_pitch.pdf}
\caption{Comparison of IR drop for different bump pitches in the 3D IC chip-on-VRM configuration}
\label{fig:bump_pitch}
\end{figure}

In this study, bump pitch is held at 100~$\mu$m for the 3D IC chip-on-VRM case. In this section, the impact of bump pitch variation has been investigated. The bump pitch has been varied from 100~$\mu$m to 500~$\mu$m. For each of the configurations, as the bump pitch increases, bump diameter is increased with the same factor. The number of TSVs under each bump is also increased in quadratic progression. With the increased pitch, the overall number of bumps will decrease. Therefore, each bump will carry more current. If the number of TSVs is not increased in quadratic progression, then the amount of current carried by each TSV will increase, leading to increased joule heating \cite{ucla:reliability} and potentially reducing the meantime to failure (MTTF)\cite{black:mttf}. Fig. \ref{fig:bump_pitch} summarizes the results. Only the 3D IC case has been considered here for the analysis as we believe other cases will follow the same trend. For larger bump pitches, since the overall number of the TSVs is the same for all the cases, and the bump resistance is decreased with increased diameter, the on-die loss is the only differentiating factor.
\fi

\section{Comparison of Transient Noise for different configurations}
\begin{figure}[!ht]
        \centering
        \includegraphics[width=0.35\textwidth]{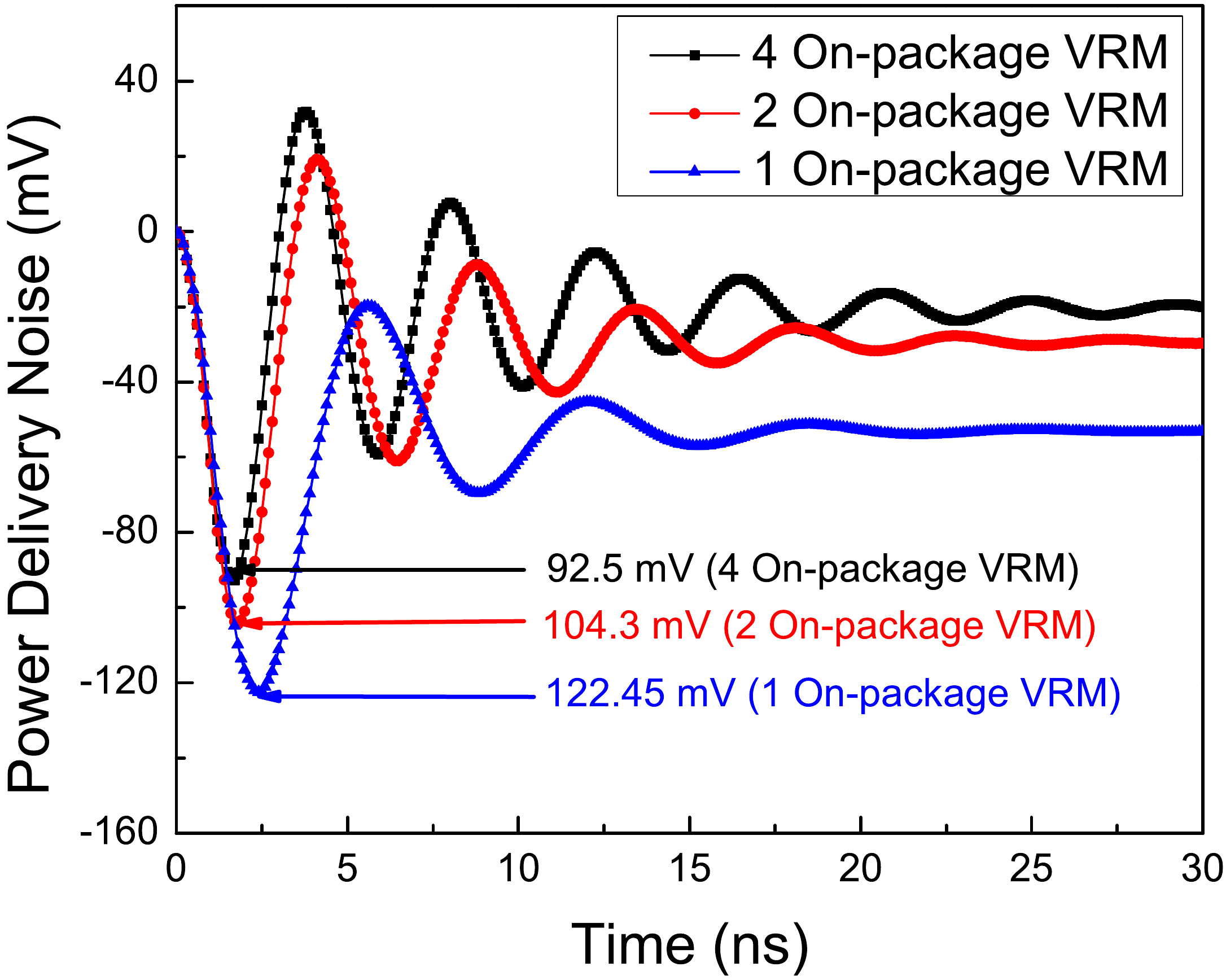}
        
    \caption{Comparison of transient noise for different on-package VRM configurations}
    \label{fig:multi_VRM}
\end{figure}

\begin{figure}[!ht]
\centering \includegraphics[width=0.35\textwidth]{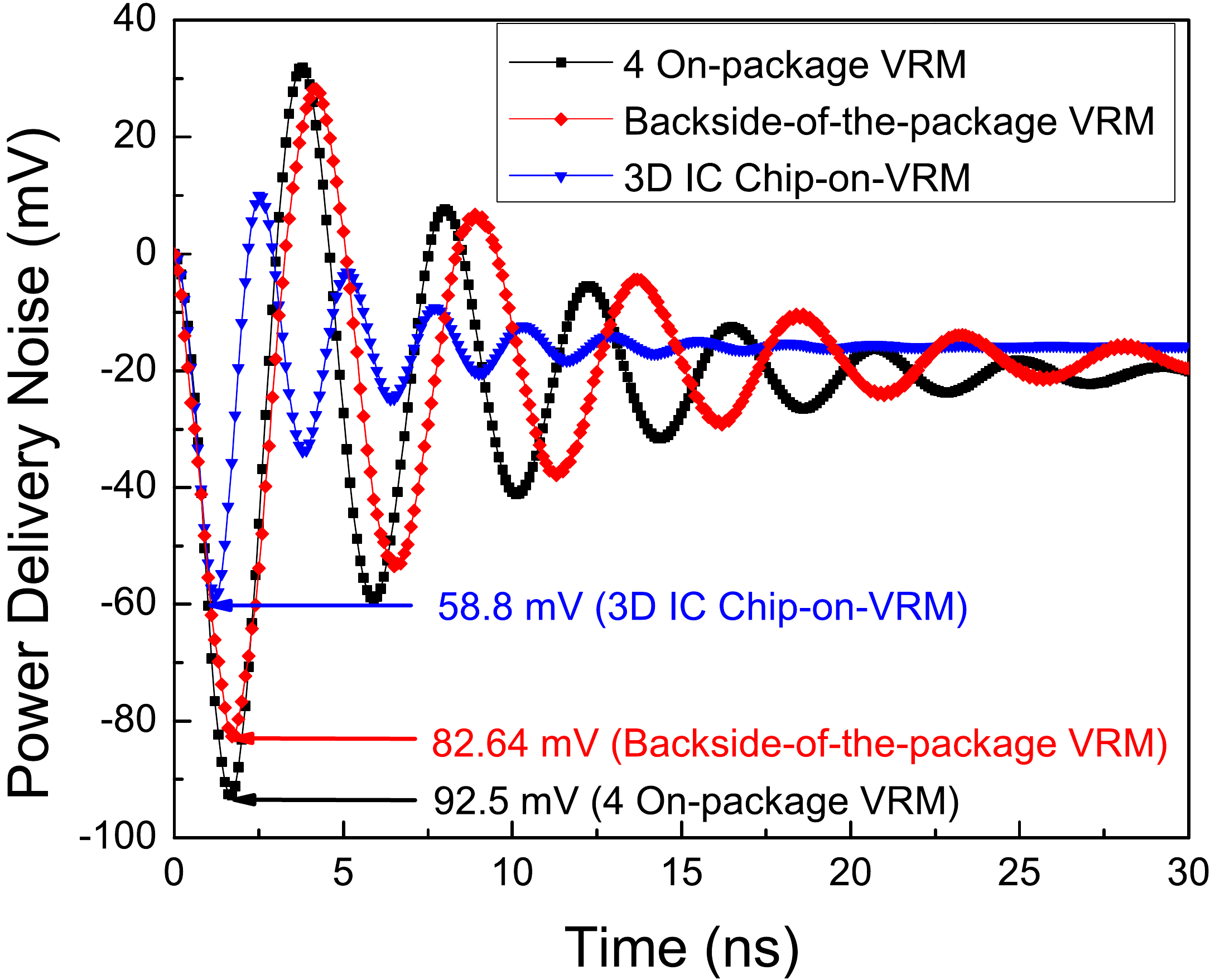}
\caption{PSN comparison for different key benchmarks}
\label{fig:multi_bench}
\end{figure}

The PDN of a system typically contains many inductive elements. Evaluating the transient i.e., $L\frac{di}{dt}$ noise of a system is thus important for verifying the noise levels. \iffalse Because of the area constraints, the amount of on-chip decoupling capacitor (decap) is very limited \cite{samsung:decap}, which affects the first droop noise of the system. But we can control the second and third droops as they are controlled by the integrated and mounted decaps in the package and the board. In all the transient simulations, the on-chip decap is fixed to the specified number as mentioned in Table \ref{tab:specs}. The controlling parameters are the discrete decaps on the board and the package. In each case, a small number of decaps is considered. Since this paper focuses on the impact of different benchmark architectures on DC IR drop and simultaneous switching noise (SSN), a detailed analysis of decap allocation for optimized result is out of the scope.\fi In this section, for different configurations, step response of the system will be shown. The supply voltage rises from 0~V to 1~V with a rise time of 1~ns. Fig. \ref{fig:multi_VRM} shows the transient noise profile for multiple on-package VRMs. As expected, with increased number of VRMs surrounding the chip, there is less PSN. In all the cases, the transient noises generated from the interaction of capacitive and inductive (mainly package) elements oscillate and settle down to the DC IR drop value of the corresponding case. The second droop is suppressed by the discrete decaps placed on the package. That is why the most dominant transient droop in all the cases is the first droop noise. The four on-package VRMs case achieves almost 24.45\% improvement in PSN compared to the single on-package VRM case.
When the VRM chip is placed on the backside of the package, VRM-to-chip PDN is mostly dominated by package vias and bumps. Package vias typically have low aspect ratio compared to TSVs; these TSVs contribute less to the resistance and more to the inductance of the system. Solder bumps between the package and the board play a similar role compared to the microbumps. Also, the number of microbumps is higher than the number of solder bumps. In the 3D IC chip-on-VRM case, the VRM is directly supplying power from the bottom of the chip. Hence, the inductive components are the TSVs in the VRM chip, and the microbumps between the VRM and the active chip. These are minimal compared to the inductive components in the other cases described in this study. In both of the cases, the package is less involved, which reduces the overall package parasitics in the PDN. Fig. \ref{fig:multi_bench} compares the best case from the on-package VRM cases with the backside-of-the-package and 3D IC chip-on-VRM configurations. For the backside-of-the-package VRM case, the maximum PSN is 82.64~mV. This itself is 10.65$\%$ improvement compared to the four on-package VRMs case. The 3D IC chip-on-VRM case provides a maximum PSN of 58.8~mV. 
 
\begin{figure}[!ht]
\centering \includegraphics[width=0.35\textwidth]{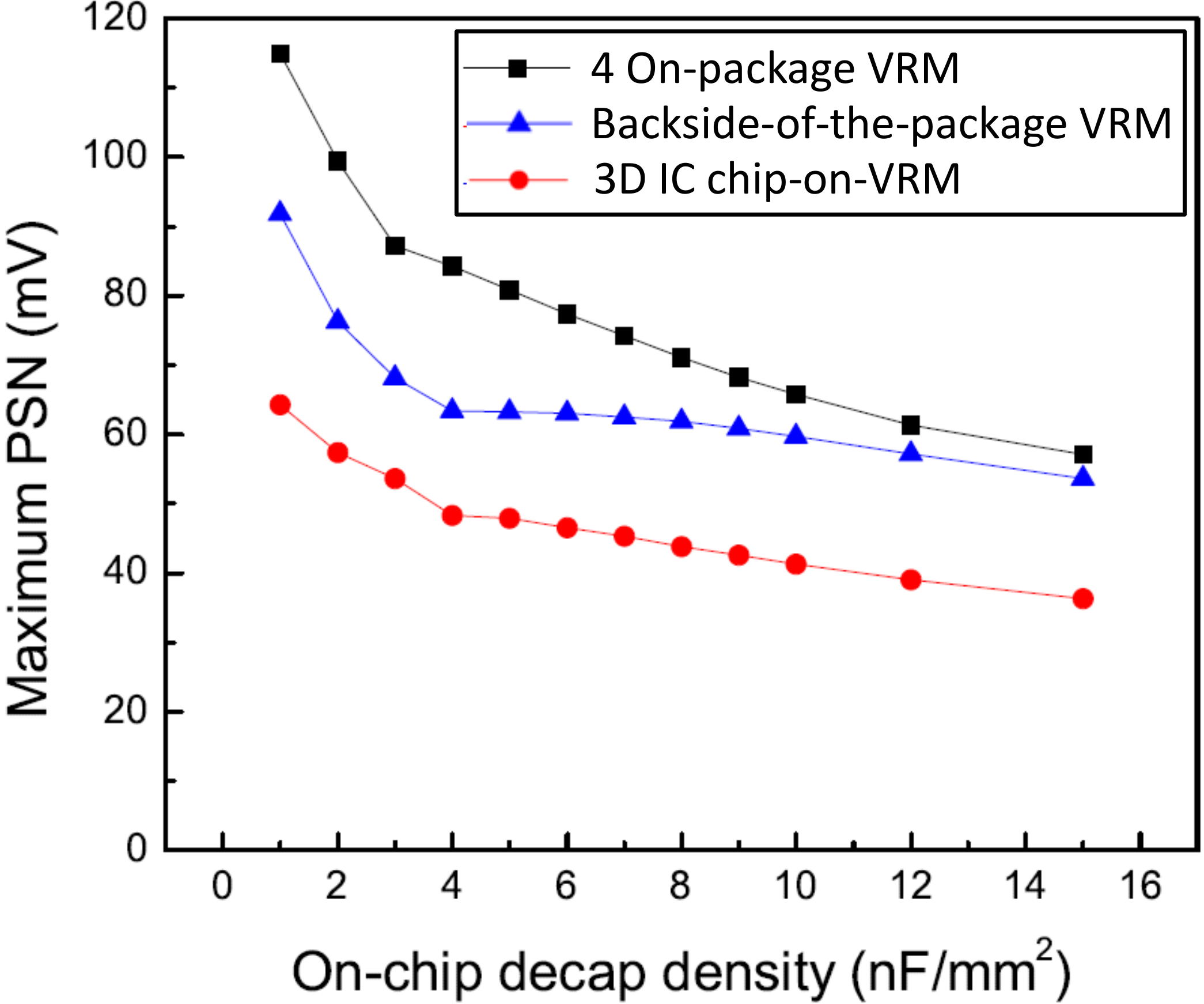}
\caption{Maximum PSN of some key configurations for different on-chip decap density}
\label{fig:onchip_decap}
\end{figure}

Throughout the paper, we observe that the transient noise is dominated by the first droop noise. This noise is dependent on the on-chip decap allocation. Throughout this paper, a decap density of 5.3 nF/mm$^2$ has been used for the analysis. Typically, on-die decap can take 20-30\% area depending on the available space \cite{samsung:decap}. Moreover, depending on the type of capacitors used, the decap density can vary \cite{kaushik_roy:decap}. Typically using MOS capacitors, a decap density of 10-20~nF/mm$^2$ can be achieved. The four on-package VRMs case, the 3D IC chip-on-VRM case, and the backside-of-the-package VRM case have been simulated for a varying decap density. The density is varied from 1~nF/mm$^2$ to 15~nF/mm$^2$. To simplify the analysis, uniform power density has been considered. Fig. \ref{fig:onchip_decap} summarizes the results from this study. As expected, with increased decap allocation, the PSN decreases. For the 3D IC case, the maximum PSN reduced from 64~mV for 1~nf/mm$^2$ to 36~mV for 15~nF/mm$^2$. The other two cases follow a trend similar to this.

\section{Conclusion}
This paper performs a power delivery network analysis for different benchmark configurations including voltage regulator modules. Multiple on-package VRMs, 3-D IC chip-on-VRM, and backside-of-the-package VRM cases are studied. The latter two cases enable supplying power directly from the bottom of the chip. Because of the proximity from the power supply to the active circuitry, the power delivery noise of the 3D IC chip-on-VRM case and the backside-of-the-package VRM case are the least. With distributed on-chip decoupling capacitor and package level discrete decaps, the PSN is minimized in all the configurations. The impact of on-chip decap density variation is also quantified. For 3D IC chip-on-VRM case with uniform current density, 25\% improvement in PSN is possible if three times more decap is used compared to the one used for this analysis.

\ifCLASSOPTIONcaptionsoff
  \newpage
\fi

\end{document}